\documentclass[reprint,
 amsmath,amssymb,
 aps,
]{revtex4-2}
\usepackage{bm, color}

\usepackage[colorlinks = true,
            allcolors  = blue]{hyperref}
\usepackage{placeins}
\usepackage{float}
\usepackage{hyperref}

\usepackage{graphicx}
\usepackage{dcolumn}
\usepackage{bm}

\usepackage{mathtools, nccmath}

\allowdisplaybreaks

\usepackage{color,soul}

\def\H{\mathcal{H}}

\begin{document}

\preprint{APS/123-QED}

\title{On the unimportance of memory for the time non-local components of the Kadanoff-Baym equations}

\author{Cian C. Reeves}
\affiliation{%
Department of Physics, University of California, Santa Barbara, Santa Barbara, CA 93117
}%
\author{Yuanran Zhu}
\author{Chao Yang}
\affiliation{Applied Mathematics and Computational Research Division, Lawrence Berkeley National Laboratory,
Berkeley, CA 94720, USA}\author{Vojt\ifmmode \check{e}\else \v{e}ch Vl\v{c}ek}
\affiliation{%
Department of Chemistry and Biochemistry, University of California, Santa Barbara, Santa Barbara, CA 93117
}%
\affiliation{%
Department of Materials, University of California, Santa Barbara, Santa Barbara, CA 93117
}

\date{\today}
\begin{abstract}
The generalized Kadanoff-Baym ansatz (GKBA) is an approximation to the Kadanoff-Baym equations (KBE), that neglects certain memory effects that contribute to the Green's function at non-equal times. Here we present arguments and numerical results to demonstrate the practical insignificance of the quantities neglected when deriving the GKBA at conditions at which KBE and GKBA are appropriate.  We provide a mathematical proof that places a scaling bound on the neglected terms, further reinforcing that these terms are typically small in comparison to terms that are kept in the GKBA.  We perform calculations in a range of models, including different system sizes and filling fractions, as well as experimentally relevant non-equilibrium excitations.  We find that both the GKBA and KBE capture the dynamics of interacting systems with moderate and even strong interactions well.  We explicitly compute terms neglected in the GKBA approximation and show, in the scenarios tested here, that they are orders of magnitude smaller than the terms that are accounted for, i.e., they offer only a small correction when included in the full Kadanoff-Baym equations.   
\end{abstract}

\maketitle

\section{Introduction}\label{sec:intro}
For a given system and set of initial conditions, the time evolution of the one particle non-equilibrium Green's function (NEGF) is governed by the Kadanoff-Baym equations(KBEs)\cite{Stan_2009}.  These are a set of coupled integro-differential equations describing the time evolution of Green's function (GF).  Memory and correlation effects are accounted for through the collision integrals. This time evolution is in principle exact, given  exact knowledge of the  self energy, which is a key component of the collision integral.  In practice, the self energy is an approximate quantity.  A major issue suffered by the KBEs is the computational cost of their propagation.  The equations must be solved at all points on a two-time grid, coupled with the calculation of the collision integrals, leading to a $O(N_t^3)$ asymptotic scaling in the number of time steps $N_t$\cite{bonitz2015quantum}. This formal time non-locality of the KBE formalism means that the propagation is often restricted to a relatively small number of time steps.

Further, this restriction often requires the KBEs to combined with application to computationally tractable problems (by means of their effective size or approximations to the self-energy). There are many interesting non-equilibrium phenomena that are observed exclusively in solid state extended systems whose energy,length, and time-scales cannot be effectively reduced, such as insulator-metal phase transitions or band inversion\cite{Beebe_2017,Hedayat_2021}.  If one is interested in studying such systems using first-principles theory it is crucial to not only be able to simulate larger system sizes but also to be able to use improved self energy approximations\cite{Golze_2019}, which can quickly make the KBEs prohibitively expensive.   While the KBEs have been applied extensively to a range of both micro and macroscopic systems\cite{Kwong_2000,Kremp_1999,Bonitz_1996,Banyai_1998,Haberland_2001,Kwong_1998,Sakkinen_2012,Dahlen_2007}, the majority of these studies are still limited to the second Born self energy with some exceptions\cite{Schuler_2020,Schlunzen_2017,Schlunzen_2016,Schlunzen_2016_2}.

To overcome some of these challenges, an approximate partial solution, known as the Hartree-Fock generalized Kadanoff-Baym ansatz(HF-GKBA), is more commonly used for simulations of larger-scale problems\cite{Lipavsky_1986,Hermanns_2012,Hermanns_2012}.  Recently the HF-GKBA has even been applied to first principles studies of extended periodic quantum systems with the $GW$ self energy\cite{Perfetto_2022}. With the HF-GKBA, only the GF at equal times is needed, and from this, the GF at non-equal times is approximately constructed.  The HF-GKBA has become widely used due to the speedup it offers over the full KBEs, especially for long-time propagation, and even mitigates errors of KBEs in reconstructing particle densities in a simple Hubbard chain\cite{Hermanns_2014}.  In its original formulation, the HF-GKBA still retains the $O(N_t^3)$ scaling (except when used with the second-Born self energy), but recently an  $O(N_t)$ reformulation of the HF-GKBA, known as the G1-G2 scheme has been introduced\cite{Joost_2020}.  This linear scaling approach gives access to more complicated self energies such as $GW$ while still being applicable to extended periodic systems, making it an important tool to study nonequilibrium solid state systems\cite{Perfetto_2022}.

The approximations made in the derivation of the HF-GKBA can be thought of as a time-off-diagonal Markovian-type approximation (i.e., a partially memory-less representation which is detailed in section \ref{sec:markovian_analysis}) of the full KBEs for the $t\neq t'$ component of the GF, since they amount to neglecting certain memory integrals in the equations of motion for the GF.  And while the linear scaling formulation of the HF-GKBA is extremely promising, it is unclear when the approximations introduced in the derivation of the HF-GKBA are justified.  In a recent study a similar type of approximation was applied\cite{Stahl_2022}. Here the range of temporal integrals is truncated to improve the computational cost. Thus showing, the system's full memory is not always needed and giving support to the approximations made in the HF-GKBA.  The quality of the ansatz was demonstrated in\cite{Kwong_1998} again hinting at the unimportance of the neglected effects.  The approximations made in deriving the HF-GKBA depend primarily on the temporal decay of many-body correlations in the system and the interaction strength.  Broadly speaking the HF-GBKA is thought to work for systems with moderate interaction strengths and allows one to study regimes beyond linear response.  Indeed, many studies show, in these regimes, the HF-GKBA can match exact results with excellent accuracy\cite{Schlunzen_2017,Bonitz_2013,Balzer_2013,Hermanns_2013}.  Furthermore, parameter regimes in which one would expect the approximations to be poor, namely strong interactions and strong excitation from equilibrium, often coincide with regimes in which even the full KBEs begin to fail.  

In this paper, we investigate the importance of the memory integrals neglected when deriving the HF-GKBA.  To see how significantly the memory affects the quality of the approximate dynamics we compare the KBEs and HF-GKBA in exactly diagonalizable models under experimentally relevant non-equilibrium preparations.  In most of our numerical tests, we find the HF-GKBA and KBE both give at least qualitatively correct results and in many cases are quantitatively correct too.  Furthermore, we find in many cases where HF-GKBA differs significantly from the exact results, so do the KBEs. For strong interactions, even sophisticated self energy approximations like $GW$ will fail to provide accurate results.  This is especially relevant for the KBEs since typically simpler and less accurate self energies than $GW$ are used.  In the case of strong non-equilibrium, the KBEs are known to suffer from artificial damping and the development of spurious steady states, which is corrected somewhat in the HF-GKBA\cite{von_Friesen_2010,von_Friesen_2009}. We explicitly compute the neglected memory integrals and compare them to the terms that are kept in the GKBA, giving numerical evidence that these terms give only a small correction to the GKBA.  Finally, as an outlook based on observations and results presented below, we argue that the use of more accurate self energy formulations is of more importance than the inclusion of the memory effects neglected in the HF-GKBA.

\section{Theory}\label{sec:Theory}

\subsection{The Kadanoff-Baym Equations}

The Kadanoff-Baym equations describe the time evolution of a two-time non-equilibrium GF initially at equilibrium and perturbed by an external field.  In this section, we will introduce the KBEs, as well as some of the theory needed to understand their meaning. 

When a system initially at thermal equilibrium at inverse temperature $\beta$ is perturbed by a time-dependent field, the expectation of an operator can be written generally as\cite{Stan_2009},
\begin{equation}
    \langle \hat{A}(t)\rangle = \frac{\textrm{Tr}[\hat{U}(-i\beta,0)\hat{U}^\dagger(0,t)\hat{A}\hat{U}(t,0)]}{\textrm{Tr}[\hat{U}(-i\beta,0)]}
\end{equation}
\FloatBarrier
\begin{figure}
    \centering
    \includegraphics[width=\linewidth]{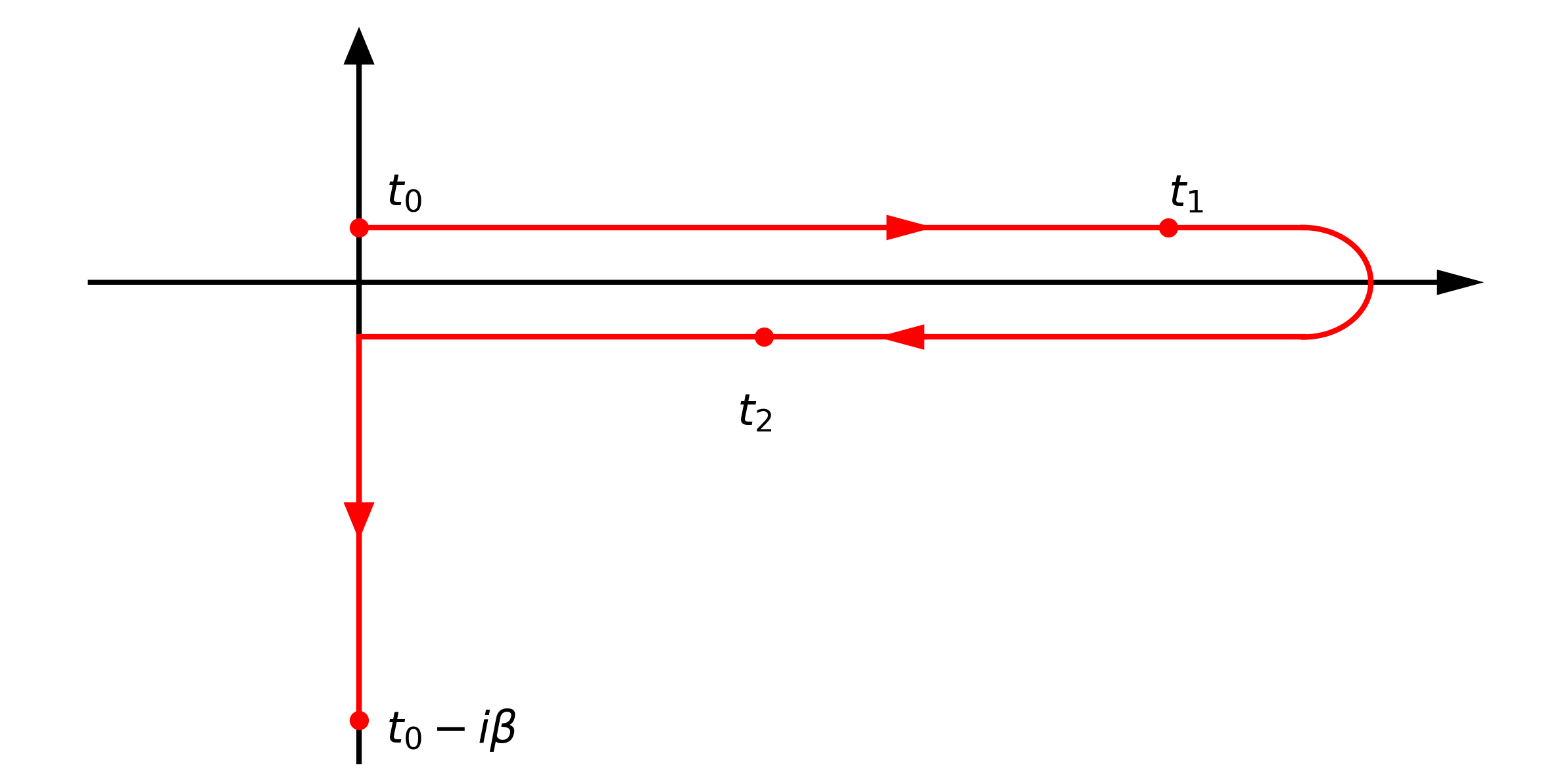}
    \caption{The Keldysh contour.  The contour time ordering operator, $\mathcal{T}_C$, places operators from right to left such that their time arguments follow the order of the arrows shown in the contour.  The vertical portion of the track relates to the initial preparation while the horizontal portion is related to the non-equilibrium evolution of the operator.}
    \label{fig:keldysh_contour}
\end{figure}
\noindent Here, without loss of generality, we assume the time-dependent field is switched on at $t=t_0=0$.  In the above we also have,
\begin{equation}
\begin{split}
    U(t_2,t_1)&=
    \begin{cases}
    \mathcal{T}\{\exp[-i\int_{t_1}^{t_2} dt\mathcal{H}(t)]\} \quad t_2 > t_1\\
    \Bar{\mathcal{T}}\{\exp[-i\int_{t_1}^{t_2} dt\mathcal{H}(t)]\} \quad t_2 < t_1,
    \end{cases}\\
\end{split}
\end{equation}
where $\mathcal{T} (\bar{\mathcal{T}})$ is the (anti-)time-ordering operator.  Finally,
\begin{equation}
        U(-i\beta,0) = \exp(-\beta \mathcal{H}(0)),
\end{equation}
is the thermal statistical averaging operator for a given inverse temperature $\beta$. In the following sections, the finite temperature formalism at a sufficiently low temperature is invoked in preparing the correlated ground state for the KBEs. In contrast, the strictly 0K formalism typically employs adiabatic connection to prepare the correlated quantum state. Within the Keldysh formalism, this product can equivalently be written as an ordered product on the contour shown in Fig. \ref{fig:keldysh_contour}.  The operator $\mathcal{T}_C$ in equation \eqref{eq:contour_G} denotes the contour ordering and places operators from right to left with time arguments in the order that corresponds to the direction of the arrows appearing in Fig. \ref{fig:keldysh_contour}. While the details of the Keldysh formalism are not crucial in this work, we do introduce some common definitions in order to present the Kadanoff-Baym equations for the NEGF. For more information on the Keldysh formalism, we direct the reader to \cite{stefanucci2013nonequilibrium,Kadanoff_1962,bonitz2015quantum,Stan_2009}. 

Firstly, in the Keldysh formalism the single particle GF can be written as,
\begin{equation}\label{eq:contour_G}
    G(t_1,t_2) = -i \langle \mathcal{T}_C[c(t_1)c^\dagger(t_2)]\rangle.
\end{equation}
At inverse temperature $\beta$ and with $t_0$ denoting the time at which the system leaves equilibrium, $t_1$ and $t_2$ can lie anywhere on the contour shown in Fig. \ref{fig:keldysh_contour}.  When both time arguments lie on the real axis the lesser and greater GFs are defined as follows,
\noindent
\begin{equation}
    \begin{split}
        G^{>}(t_1,t_2) &= -i\langle c(t_1) c^\dagger(t_2)\rangle,\\
        G^{<}(t_1,t_2) &= i\langle  c^\dagger(t_2)c(t_1)\rangle.
    \end{split}
\end{equation}
The time ordering operator means the full GF in equation \eqref{eq:contour_G} will be given by $G^{>}(t_1,t_2)$ when $t_1$ is later than $t_2$ and $G^{<}(t_1,t_2)$ when $t_2$ is later than $t_1$ and both are on the real axis.  When one of the time arguments lies on the imaginary axis we have,
\begin{equation}
\begin{split}
        G^{\lceil}(-i\tau,t) &= G^{>}(t_0-i\tau,t),\\
        G^{\rceil}(t,-i\tau) &= G^{<}(t,t_0-i\tau).\\
\end{split}
\end{equation}
Finally, when both arguments lie on the imaginary axis we are left with the time translation invariant Matsubara GF,
\begin{equation}
    iG^\mathrm{M}(\tau_1-\tau_2) = G(t_0 - i\tau_1,t_0 - i\tau_2),
\end{equation}
which represents the equilibrium GF at a given temperature. The same definitions also exist for the self energy operator.
Using these notations the KBEs for time propagation of the NEGF can be written explicitly as,
\begin{equation}\label{eq:KBE}
           \begin{split}
                [-\partial_\tau - h] G^\mathrm{M}(\tau) &= \delta(\tau) + \int_0^\beta d\Bar{\tau}\Sigma^\mathrm{M}(\tau-\Bar{\tau})G^\mathrm{M}(\bar{\tau}),\\
                i\partial_{t_1} G^{\rceil}(t_1,-i\tau) &= h^{\textrm{HF}}(t_1)G^{\rceil}(t_1,-i\tau) + I^{\rceil}(t_1,-i\tau),\\
                -i\partial_{t_2} G^{\lceil}(-i\tau,t_2) &= G^{\lceil}(-i\tau,t_2)h^{\textrm{HF}}(t_2) + I^{\lceil}(-i\tau,t_1),\\
                i\partial_{t_1} G^{\lessgtr}(t_1,t_2) &= h^{\textrm{HF}}(t)G^{\lessgtr}(t_1,t_2) + I_1^{\lessgtr}(t_1,t_2),\\
                -i\partial_{t_2} G^{\lessgtr}(t_1,t_2) &= G^{\lessgtr}(t_1,t_2)h^{\textrm{HF}}(t_2) + I_2^{\lessgtr}(t_1,t_2),\\
            \end{split}
\end{equation}
with the so-called collision integrals being given by

\begin{equation}\label{eq:coll_int}
    \begin{split}
        I_{1}^{\lessgtr}(t_1,t_2) &= \int_{0}^{t_1}\mathrm{d}\bar{t} \Sigma^\mathrm{R}(t_1,\Bar{t})G^{\lessgtr}(\Bar{t},t_2) \\ &\hspace{10mm}+\int_{0}^{t_2} \mathrm{d}\bar{t} \Sigma^{\lessgtr}(t_1,\Bar{t})G^\mathrm{A}(\Bar{t},t_2)\\ &\hspace{15mm}- i\int_0^\beta \mathrm{d}\bar{\tau} \Sigma^\rceil(t_1,-i\bar{\tau})G^{\lceil}(-i\bar{\tau},t_2)\\
        I_{2}^{\lessgtr}(t_1,t_2) &= \int_{0}^{t_1} \mathrm{d}\bar{t} G^\mathrm{R}(t_1,\Bar{t})\Sigma^{\lessgtr}(\Bar{t},t_2) \\&\hspace{10mm}+ \int_{0}^{t_2} \mathrm{d}\bar{t} G^{\lessgtr}(t_1,\Bar{t})\Sigma^\mathrm{A}(\Bar{t},t_2) \\&\hspace{15mm}- i\int_0^{\beta}\mathrm{d}\bar{\tau}G^{\rceil}(t_1,-i\bar{\tau})\Sigma^{\lceil}(-i\bar{\tau},t_2)\\
        I^{\rceil}(t_1,-i\tau) &= \int_{0}^{t_1} d\bar{t} \Sigma^\mathrm{R}(t_1,\bar{t})G^{\rceil}(\bar{t},-i\tau)\\&\hspace{15mm} + \int_0^\beta d\bar{\tau} \Sigma^{\rceil}(t_1,-i\bar{\tau})G^M(\bar{\tau} - \tau),\\
        I^{\lceil}(-i\tau,t_1) &= \int_0^{t_1}d\bar{t} G^{\lceil}(-i\tau,\bar{t})\Sigma^\mathrm{A}(\bar{t},t) \\ &\hspace{15mm}+\int_0^{\beta}d\bar{\tau}G^\mathrm{M}(\tau - \bar{\tau}) \Sigma^{\lceil}(-i\bar{\tau},t_1).
    \end{split}
\end{equation}
Here the retarded/advanced Green's function $G^{\mathrm{R/A}}$ and self energy $\Sigma^{\mathrm{R/A}}$ are functions of $G^{\lessgtr}$. In the above equations, the self energy $\Sigma(t_1,t_2)$ includes only correlation terms, while the Hartree-Fock contribution is included in $h^{\mathrm{HF}}(t)$. The first equation in \eqref{eq:KBE} describes the role of the initial correlations in the propagation of the NEGF. The remaining equations describe the propagation of the two-time particle and hole propagator after leaving equilibrium. 

The two-time nature of the KBEs combined with the various collision integrals means the cost of solving these equations scales cubically in the number of time steps.  Several approaches have been employed to circumvent the difficulty of performing long NEGF time evolutions.  This includes extrapolation of trajectories from a short snapshot of the initial dynamics as well reducing cost through stochastic compression of matrix contractions\cite{Reeves_2023,Yin_2021,mejía_2023} . Another approach is through direct approximation of the full KBEs.  The most popular of these approximation schemes is known as the Hartree-Fock generalized Kadanoff-Baym ansatz (HF-GKBA).  In the following section, we will introduce the HF-GKBA.

\subsection{The Hartree-Fock Generalized Kadanoff-Baym Ansatz}\label{sub_sect:HFGKBA}
Unlike in the KBEs in the HF-GKBA initial correlations are not prepared through the contour integration but rather through other means such as adiabatic switching\cite{Touvinen_2019,Karlsson_2018}. For this reason only the KBEs with real time arguments are considered in the derivation of the GKBA.  The final two equations in equation \eqref{eq:KBE} can be combined to give a single equation for the time diagonal GF,
\begin{equation}\label{eq:time_diagonal}
           \begin{split}
                i\partial_t G^{<}(t,t) &= [h^{\textrm{HF}}(t),G^{<}(t,t)] + I_1^<(t,t) - I_2^{<}(t,t)
            \end{split}
        \end{equation}
with 
\begin{equation}\label{eq:coll_int_GKBA}
    \begin{split}
        I_{1}^{\lessgtr}(t_1,t_2) &= \int_{0}^{t_1} \mathrm{d}\bar{t} \Sigma^\mathrm{R}(t_1,\Bar{t})G^{\lessgtr}(\Bar{t},t_2) \\ &\hspace{20mm}+\int_{0}^{t_2} \mathrm{d}\bar{t} \Sigma^{\lessgtr}(t_1,\Bar{t})G^\mathrm{A}(\Bar{t},t_2)\\
        I_{2}^{\lessgtr}(t_1,t_2) &= \int_{0}^{t_1} \mathrm{d}\bar{t} G^\mathrm{R}(t_1,\Bar{t})\Sigma^{\lessgtr}(\Bar{t},t_2) \\&\hspace{20mm}+ \int_{0}^{t_2} \mathrm{d}\bar{t} G^{\lessgtr}(t_1,\Bar{t})\Sigma^\mathrm{A}(\Bar{t},t_2).
    \end{split}
\end{equation}
This form of the collision integrals assumes the state at $t=0$ has already been prepared in the correlated ground state.  The preparation via adiabatic switching will be discussed in section \ref{sec:Methods}.  We note that now the time arguments lie strictly on the real-time axis. 

The HF-GKBA is derived directly from the KBE and can be summarized in the following equations\cite{Hermanns_2012},
\begin{equation}\label{eq:HF-GKBA}
   \begin{split}
        G^{\lessgtr}(t_1,t_2) &= iG^\mathrm{R}(t_1,t_2)G^{\lessgtr}(t_2,t_2) - iG^{\lessgtr}(t_1,t_1)G^\mathrm{A}(t_1,t_2),\\
        G^{\mathrm{R,A}}(t_1,t_2)&=\pm \Theta[\pm(t_1 - t_2)]T\{\mathrm{e}^{-i\int_{t_2}^{t_1} h^{\textrm{HF}}(t) dt}\}.
   \end{split}
\end{equation}
In other words, at each time step only equation \eqref{eq:time_diagonal} is explicitly evaluated. Equation \eqref{eq:HF-GKBA} is then used to reconstruct the time off-diagonal components.

Apart from those approximations made to the self energy, which HF-GKBA and KBE share,  two additional approximations are made in the derivation of HF-GKBA.  The first involves neglecting integrals discussed in detail below and given explicitly for the lesser Green's function in equation \eqref{eq:G_lss_full}.  These terms appear in the expression for reconstructing $G^{\lessgtr}(t,t')$.  Once dropped, one is left with the first expression in equation \eqref{eq:HF-GKBA}.  With no further approximation, this ansatz for the time off-diagonal components of $G^{\lessgtr}(t,t')$ is referred to as the generalized Kadanoff-Baym ansatz(GKBA)\cite{Lipavsky_1986}.   This part of approximation will be discussed in more detail in section \ref{sec:markovian_analysis}. The HF-GKBA involves a further approximation where the full $G^{\mathrm{R/A}}(t,t')$ are replaced by the retarded and advanced Hartree-Fock propagator.   Notably, the HF-GKBA leaves important quantities such as energy and particle number conserved as well as retaining causal time evolution.

Recently a linear time scaling$[\sim O(N_t)]$ implementation of the HF-GKBA has been achieved, opening the door for long-time evolution of NEGFs\cite{Joost_2020}. The method removes the explicit appearance of integrals in equation \eqref{eq:coll_int_GKBA} from the differential equation for $G^<(t)$ by expressing them in terms of the correlated part of the equal time two-particle GF $\mathcal{G}(t)$. Within this formulation $\mathcal{G}(t)$ is propagated simultaneously with $G^<(t)$ using an expression analogous to equation \eqref{eq:time_diagonal}.  Throughout this paper, we use this propagation scheme to generate HF-GKBA results for the models discussed in section \ref{sec:model_systems}.  

The exact equation of motion for $\mathcal{G}(t)$ depends on the self energy approximation used. Here, we use the second Born (SB) approximation which is commonly used in  KBEs for practical reasons\cite{Balzer_2010,Dahlen_2007}, though several works in recent years also employed more advanced self energy formulations \cite{Schuler_2020,Schlunzen_2017,Schlunzen_2016,Schlunzen_2016_2}. The computation of the SB is relatively quick compared to other more complicated expressions like $GW$, thus does not add much overhead to the propagation.

For the SB self energy, the equations of motion for $G^<(t)$ and $\mathcal{G}(t)$ in the orbital basis are given below\cite{Schlunzen_2020}.
\begin{equation}\label{eq:G1-G2}
\begin{split}
       i \partial_t G^<_{ij}(t) &= [h^{\textrm{HF}}(t), G^<(t)]_{ij} + [I+I^\dagger]_{ij}(t)\\   i\partial_t \mathcal{G}_{ijkl}(t) &= [h^{(2),\textrm{HF}}(t),\mathcal{G}(t)]_{ijkl} +\Psi_{ijkl}(t).
\end{split} 
\end{equation}
Above, the following definitions are made,
\begin{align}
        h_{ij}^{\textrm{HF}}(t) &= h^{(0)}_{ij}(t) - i\sum_{kl} [2w_{ikjl}(t) - w_{iklj}(t)]G_{kl}^<(t)\nonumber,\\
        I_{ij}(t)&=-i\sum_{klp} w_{iklp}(t)\mathcal{G}_{lpjk}(t),\nonumber\\
        h^{(2),\textrm{HF}}_{ijkl}(t) &= \delta_{jl}h^{\textrm{HF}}_{ik}(t) + \delta_{ik}h^{\textrm{HF}}_{jl}(t)\nonumber,\\  
    \Psi_{ijkl} &=\sum_{pqrs}[w_{pqrs}(t) - w_{pqsr}(t)]\times\nonumber\\ \times\bigg[G^>_{ip}(t) &G^<_{rk}(t) G^>_{jq}(t)G^<_{sl}(t)- G^<_{ip}(t)G^>_{rk}(t)G^<_{jq}(t)G^>_{sl}(t)\bigg].
\end{align}
Here $h^{(0)}(t)$ is the single particle Hamiltonian and $w_{ijkl}(t)$ is the two-body interaction matrix.   The time dependence given to $w_{ijkl}$ is to allow for adiabatic switching for preparation of the initial state. The explicit form of both these terms will be given in section \ref{sec:model_systems}.  The tensor $\Psi_{ijkl}$ accounts for pair correlations built up due to two-particle scattering events\cite{Joost_2020}.

\subsection{Analysis of the GKBA approximation}\label{sec:markovian_analysis}
In this section, we will analyze the approximation that leads to the GKBA. The exact expression for the off-diagonal lesser GF is~\cite{Hermanns_2014},
\begin{widetext}
\begin{align}\label{eq:G_lss_full}
    G^{<}(t_1,t_2) &= \underbrace{iG^{\textrm{R}}(t_1,t_2) G^{<}(t_2,t_2) + \overbrace{\int_{t_2}^{t_1}\mathrm{d}t\int_{t_0}^{t_2}\mathrm{d}t'G^{\mathrm{R}}(t_1,t)\Sigma^{<}(t,t')G^{\mathrm{A}}(t',t_2) + \int_{t_2}^{t_1}\mathrm{d}t\int_{t_0}^{t_2}\mathrm{d}t'G^{\mathrm{R}}(t_1,t)\Sigma^{\mathrm{R}}(t,t')G^{<}(t',t_2)}^{\textrm{Memory integrals}}}_{t_1>t_2}\nonumber\\
    &-\underbrace{iG^{<}(t_1,t_1)G^{\mathrm{A}}(t_1,t_2) + \overbrace{\int_{t_2}^{t_1}\mathrm{d}t\int_{t_0}^{t_2}\mathrm{d}t'G^{\mathrm{R}}(t_1,t)\Sigma^{<}(t,t')G^{\mathrm{A}}(t',t_2) +\int_{t_2}^{t_1}\mathrm{d}t\int_{t_0}^{t_2}\mathrm{d}t'G^{<}(t_1,t)\Sigma^{\mathrm{A}}(t,t')G^{\mathrm{A}}(t',t_2)}^{\textrm{Memory integrals}} }_{t_1<t_2},
\end{align}
\end{widetext}
with a similar expression existing for the greater component of the GF. To derive the GKBA equations, the integral terms appearing in equation \eqref{eq:G_lss_full} are dropped. This level of approximation leaves us with the first line in equation \eqref{eq:HF-GKBA}.  It is important to note that the GKBA still retains \textit{some} memory effects through the collision integrals in equation \eqref{eq:coll_int_GKBA} and so is \textit{not} fully Markovian equation.  The quality of this approximation depends on several factors, including the magnitude of $w_{ijkl}$ and the type of non-equilibrium excitation.  In the extreme limit where $\Sigma(t,t')\sim \delta(t- t')$, the integral terms vanish and the GKBA becomes an exact expression.  This corresponds to the case where the system is completely memoryless, and so we should expect the approximation to be exact.

Suppose now we relax the condition, so $\Sigma(t,t')\nsim \delta(t- t')$,  but still assume $\Sigma(t,t')$ to have a decay envelope and decay over a timescale set by some time parameter $\tau_c$.  The quantity $\tau_c$ can be interpreted as the time over which many-body correlations decay. Because the GFs in equation \eqref{eq:G_lss_full} are bounded, the integrand will also decay on a similar timescale $\tau_c$.  We can now imagine splitting the domain of integration over $t$ and $t'$ into a region with $|t-t'|<\tau_c$ and a complementary region where $|t-t'|>\tau_c$ Due to the assumed decay of the self energy, the integral over the latter region will be negligible.  For the integrals in equation \eqref{eq:G_lss_full} to provide a significant correction to the GKBA the time $\tau_c$ must be long enough to allow the memory integrals to build up to be on the order of the non-integral terms.

As is mentioned in the original paper on the GKBA\cite{Lipavsky_1986}, for temporal separations less than $\tau_c$, exact statements about the integral are difficult to make. This is because they depend more heavily on the structure of the non-decaying parts of the self energy and GF.  We have however observed numerically that in many circumstances, the self energy, and more specifically the integrands appearing in equation \eqref{eq:G_lss_full} are relatively small in magnitude, when comparing with the Green's function.  Results to demonstrate this will be given and discussed in section \ref{sec:numerical_results}. Additional results are also given in the supplementary materials~\cite{supp}.  One major reason for this is that the self energy and thus the integral terms are  typically proportional to the interaction strength squared. For example, with the second Born self energy used here we have,
\begin{align}\label{2ndB_selfE}
    \Sigma^{\lessgtr}_\mathrm{SB}(t,t')\sim w^2 G^\lessgtr(t,t') G^\lessgtr(t,t') G^\gtrless(t',t),
\end{align}
where $w$ represents to the two-body interaction matrix.  Similarly, to leading order, the $GW$ and T-matrix self energies scale quadratically with the magnitude of $w_{ijkl}$ \cite{Joost_2020}. This means that, in systems with moderate coupling strengths, the integrals appearing in equation \eqref{eq:G_lss_full} can be much smaller than the non integral terms that appear in the same expression, which is demonstrated numerically in panel B of Fig. \ref{fig:Dipole_kbe_vs_gkba} and Fig. \ref{fig:Xray_kbe_vs_gkba}. Furthermore, looking at the equation \eqref{eq:time_diagonal} for the time diagonal component, the time off-diagonals enter only through the collision integrals, and so are multiplied by another factor of the interaction strength squared.  In other words, the integral terms in equation \eqref{eq:G_lss_full} only enter the equation for the time diagonal at fourth order in the coupling.  

Another contributing factor to the small magnitude of the self energy and integrands in \eqref{eq:G_lss_full} is the uniform boundedness $|G_{ij}^{\lessgtr}(t,t')|\leq c$, where $0<c\leq1$. For the second Born self energy \eqref{2ndB_selfE}, this eventually leads to $O(w^2c^3)$ scaling of the magnitude. The detailed mathematical analysis for this scaling limit is provided in Appendices \ref{sec:app_selfE} and \ref{sec:app_selfE_NE}. The $c^3$ factor explains the magnitude differences between the integrands and  Green's function as shown in panel D of Fig. \ref{fig:Dipole_kbe_vs_gkba} and Fig. \ref{fig:Xray_kbe_vs_gkba} where $w=U=1$. We note that while in principle $c$ can equal one this occurrence is rare and it is typically strictly less than one, more details can be found in Appendices \ref{sec:app_selfE} and \ref{sec:app_selfE_NE}.  The $GW$ approximation is another commonly used self energy approximation in Green's function theory and is the workhorse for electronic structure calculations in in solids.  Due to its relevance we also consider the scaling of this term.  If we assume convergence of the Dyson summation, the leading order term in the self energy will have the same form as equation \eqref{2ndB_selfE} and so the same arguments will hold for it's scaling

The typically small integrand means that in order to have a relevant correction to the GKBA build up, not only must the self energy decay slowly away from the time diagonals but also the time evolution must be sufficiently long. We refer to this as memory build up and based on numerical observations we argue that in many practically relevant cases for which an approximate self energy captures well the ground state and for which the KBEs perform well in capturing dynamics of the density matrix, this memory build up has a negligible contribution. Thus the GKBA yields a similar quality of result with much less computational cost.

There are exceptions to these arguments: for example in systems in which there are strong interactions, the self energy will no longer necessarily be scaled by a small prefactor.  Indeed, in the case of strong interactions a more pronounced difference between the KBEs and the HF-GKBA is observed in our numerical studies. However, in strongly interacting systems many-body perturbation theory (MBPT) methods begin to break down already in equilibrium and so it is unreasonable to think non-equilibrium extensions of these methods would fair any better in similar regimes.  So, while in our numerical tests we do observe that, with strong interactions, both methods deviate significantly from one another. We also observe that both methods deviate significantly from exact results too.  In these parameter regimes, either improved self energies\cite{Mejuto_2022} or a more sophisticated explicitly correlated schemes will be required to produce more reliable results.


Another case where these arguments may no longer hold is in the case of long-lived many-body correlations. This will cause the decay of the self energy on the off-diagonals to be slowed to allow for the aforementioned memory build up to occur and lead to more significant differences between the two methods. However, we note that the time evolution will still need to be sufficiently long to allow the integrals to build up in magnitude and still may not have an effect on timescales of interest.

Finally, the scaling bound of the self energy derived in Appendices \ref{sec:app_selfE} and \ref{sec:app_selfE_NE} depends on the interaction strength as well as the diagonal elements of $G^{\gtrless}(t)$.  These entries are in turn related to the filling fraction, in general the average magnitude of these terms can be larger for fillings away from half-filling.  For this reason it may be that in systems with very high or low fillings the neglected integrals may become large enough to significantly effect the quality of the GKBA.  In section \ref{sec:further_analysis} we investigate the effect of different filling factors on the GKBA.

Based on the above we believe that in a wide range of interacting systems, in regimes where equilibrium MBPT is a reasonable theory, the KBEs and HF-GKBA will give a similar description of the nonequilibrium dynamics of the system.

In section \ref{sec:numerical_results} we provide numerical evidence for our arguments in systems characterized by long-range Coulomb interactions and different types of non-equilibrium preparations.  We look both at the deviations of the KBEs and HF-GKBA from one another and from exact diagonalization results.  We also explicitly calculate the integral terms appearing in \eqref{eq:G_lss_full} to examine the relationship between their magnitude and deviations between the HF-GKBA and the full KBEs.

\section{Methods}\label{sec:Methods}
To demonstrate the arguments outlined in section \ref{sec:markovian_analysis}, we perform simulations for a model after excitation by different forms of non-equilibrium perturbation.  We benchmark against exact diagonalization to see how and when the KBEs improves upon HF-GKBA.  The specifics of the models we study are outlined in section \ref{sec:model_systems}, but here we give relevant information about the parameters used in this paper.

In the HF-GKBA simulations we use a code we have developed that is also publicly available online\footnote{\url{https://github.com/VlcekGroup/G12KBA.git}}.  Equation \eqref{eq:G1-G2} was propagated using fourth-order Runge-Kutta with a time step of $0.02 J^{-1}$.  We solve the KBEs using the NESSi library\cite{Schuler_2020}.  For these results, we employ a slightly larger time step of $0.025J^{-1}$.  For simulating zero-temperature dynamics, in the KBE solver, we increase the inverse temperature $\beta$ until the dynamics no longer changed.  We use $\beta = 40$ in this work.

We performed calculations in an interacting lattice model with Coulomb interactions scaled by $\gamma = 0.5$, and given explicitly in equation \eqref{eq:two-body} in the next section, that details the model Hamiltonian. To increase the generality of our analysis,  we study models with both nearest neighbor and long-range hopping between sites.  The hopping term is assumed to decay exponentially with respect to 
 distance with a decay parameter $\alpha$.  We take $\alpha=\infty$, for nearest neighbor hopping, and $\alpha=0.7$, for long-range hopping. For each system setup, calculations were run for $U = 0.1J$, $0.2J$, $0.5J$, $1.0J$ and $1.5J$.  

The two ways we drive the system out of equilibrium are described in section \ref{sec:model_systems}.  The Hamiltonian for each is given by equations \eqref{eq:dipole} and \eqref{eq:X-ray}.  Both are characterized by three parameters.  The magnitude is determined by a variable $E$, which we take to be $0.5J$ and $1.0J$.   The width and midpoint of the perturbation is determined by $\sigma$ and $t_0$ respectively.  These values remain fixed for all the simulations and have values $\sigma=0.5J^{-1}$ and $t_0=5J^{-1}$.  Various values of $\sigma$ where investigated and $\sigma = 0.5J^{-1}$ was chosen as it seemed to give rise to the most interesting dynamics.  Sample results showing the exact and GKBA dynamics for different $\sigma$ are shown in Fig. S11 of the supplementary\cite{supp}.

For all three propagation methods we prepare the system in the respective correlated initial state before initiating the excitation from equilibrium.   In the case of exact diagonalization, we can trivially prepare the system in the exact ground state, by diagonalizing and finding the eigenstate with the lowest energy.  For the HF-GKBA we start by preparing the system in the non-interacting ground state and then time evolving with equation \eqref{eq:G1-G2} while slowly turning on the $U$ parameter.  We choose the Fermi-Dirac distribution function as our switching function,
\begin{equation}
    f_\tau(t-t_0) = \frac{1}{1+\exp\left(-\frac{t-t_0}{\tau}\right)}.
\end{equation}
We found the values $t_0 = 25 $ and $\tau = 3$ gave a sufficiently slow rate of switching to converge the models and parameters presented here.  For the KBEs the initial state is prepared by firstly solving the Matsubara equation of motion and then propagating the GF with the equations in \eqref{eq:KBE}

\section{Numerical Results}\label{sec:numerical_results}
In this section, we provide and analyze numerical results to demonstrate the arguments put forth in section \ref{sec:markovian_analysis}. In section \ref{sec:model_systems} we introduce the systems we investigate, in sections \ref{sec:long_wavelength} and \ref{sec:short_wavelength} we provide results for the two types of non-equilibrium excitation we study and in section \ref{sec:further_analysis} we provide further results for a larger model system and different filling fractions.  For observables related to the time diagonal GF we demonstrate that in many of our test models the improvement the KBEs give over the HF-GKBA is marginal. We also explicitly compute the memory integrals in equation \eqref{eq:G_lss_full} and compare them to the non integral terms in the same equation.   We use the time-dependent density as a point of comparison between the HF-GKBA and the KBEs.  This is because time-dependent density is an important quantity and can be used to derive other observables such as the dipole, and so is a good figure of merit for any time-dependent method.  We will start this section by describing the model systems we will use for our investigation.

\subsection{Model systems}\label{sec:model_systems}

To illustrate the arguments put forth in section \ref{sec:markovian_analysis}, we compare the nonequilibrium dynamics in various systems which will now be described. The parameters for these models can be found in section \ref{sec:Methods}. A generic many-body Hamiltonian can be written in the following form
\begin{equation}\label{eq:MB_ham}
    \mathcal{H}  = \sum_{ij}h^{(0)}_{ij}(t)c^\dagger_ic_j + \frac{1}{2}\sum_{ijkl} w_{ijkl}(t) c^\dagger_ic_j^\dagger c_l c_k.
\end{equation}
Here $w_{ijkl}$ is the bare two-body interaction term and $h^{(0)}(t)$ is the single-particle Hamiltonian.  We take $w_{ijkl}$ to be a density-density interaction with a $\frac{1}{r}$ decay, explicitly given by

\begin{equation}\label{eq:two-body}
    \begin{split}
        w_{ijkl}^{\sigma_i\sigma_j\sigma_k\sigma_l}(t) &= U(t)\delta_{ij}\delta_{ik}\delta_{il}\delta_{\sigma_i\sigma_k}\delta_{\sigma_j\sigma_l}(1-\delta_{\sigma_i\sigma_j}) \\&\hspace{5mm}+ U(t)\sum_{n=1}^{N_s}\frac{\gamma}{|i - j|}\delta^{(n)}_{ij}\delta_{ik}\delta_{jl}\delta_{\sigma_i\sigma_k}\delta_{\sigma_j\sigma_l},
    \end{split}
\end{equation}
where the $\sigma \in \{\uparrow,\downarrow\}$ are spin indices, $N_s$ is the number of sites and $\gamma$ determines magnitude of the long range interactions. Here we define $\delta^{(n)}_{ij}$ to be non-zero only if $|i-j|=n$.  As described in section \ref{sec:Methods}, when using the HF-GKBA, we prepare correlated initial states with adiabatic switching.  For this reason, we include an explicit time-dependence in the interaction terms above.   

The single particle Hamiltonian for our model is given by
\begin{equation}\label{eq:single_particle}
    \begin{split}
        h^{(0)}_{ij}(t) &= J e^{-\alpha(|i - j|-1)}(1-\delta_{ij}) + h_{ij}^{\mathrm{N.E}}(t)
    \end{split}
\end{equation}
The first term corresponds to the kinetic energy or hopping term between different sites of the chain. Here, in addition to long-range interactions, we allow for long-range hopping in order to improve the generality of our investigation.  The second term in equation \eqref{eq:single_particle} couples the system to a time-dependent external field, which kicks the system from equilibrium. In this paper, we take two distinct forms of excitation which will now be described.  

For each type of non-equilibrium preparation, we define the following variables.  Firstly, $E$ is a scaling factor that is related to the amplitude of the excitation, discussed already in the preceding section. We also define the lattice coordinates $r_i$ for the chain as,
\begin{equation}\label{eq:lattice_coords}
    r_i = \frac{N}{2} - i - 0.5 \quad i = 0,\dots, N-1.
\end{equation}
\noindent 
Finally, in the following expressions, $t_0$ is related to the time at which the system is driven from equilibrium, and $\sigma$ determines the temporal width of the excitation field. 

Many non-equilibrium experiments involve excitation of the system via coupling to a time-dependent electric field.  To illustrate the arguments in experimentally relevant setups,  we choose two distinct forms of $h_{ij}^{\mathrm{N.E}}(t)$ that will mimic coupling to electric fields in different parts of the electromagnetic spectrum.  The first form of excitation is taken to be within the dipole approximation, where we assume the electric field is incident in the direction along the chain. In this case $h_{ij}^{\mathrm{N.E}}(t)$ is given explicitly by
 \begin{equation}\label{eq:dipole}
     h_{ij}^{\mathrm{N.E}}(t) = \delta_{ij}r_iEe^{-\frac{(t-t_0)^2}{2\sigma^2}}.
 \end{equation}
 This can be thought to represent the long wavelength limit of an incident electromagnetic pulse.

For the second form of $h_{ij}^{\mathrm{N.E}}(t)$ we again use a Gaussian profile for the time-dependent pulse on each site. However, instead of the dipole approximation we choose,
\begin{equation}\label{eq:X-ray}
    h_{ij}^{\mathrm{N.E}}(t) = -\delta_{ij}\cos\left(\frac{\pi r_i}{2}\right)Ee^{-\frac{(t-t_0)^2}{2\sigma^2}}.
 \end{equation}
This setup is chosen to resemble a short wavelength standing wave that is ramped up and back down with a Gaussian profile. In this setup, the wavelength of the excitation is on the order of the lattice spacing.  This is the case for example when comparing typical X-ray wavelengths and lattice spacings.  
We will now analyze a selected set of the results generated in this investigation.

\subsection{Long wavelength limit}\label{sec:long_wavelength}

\begin{figure*}
    \centering
    \includegraphics[width=\textwidth]{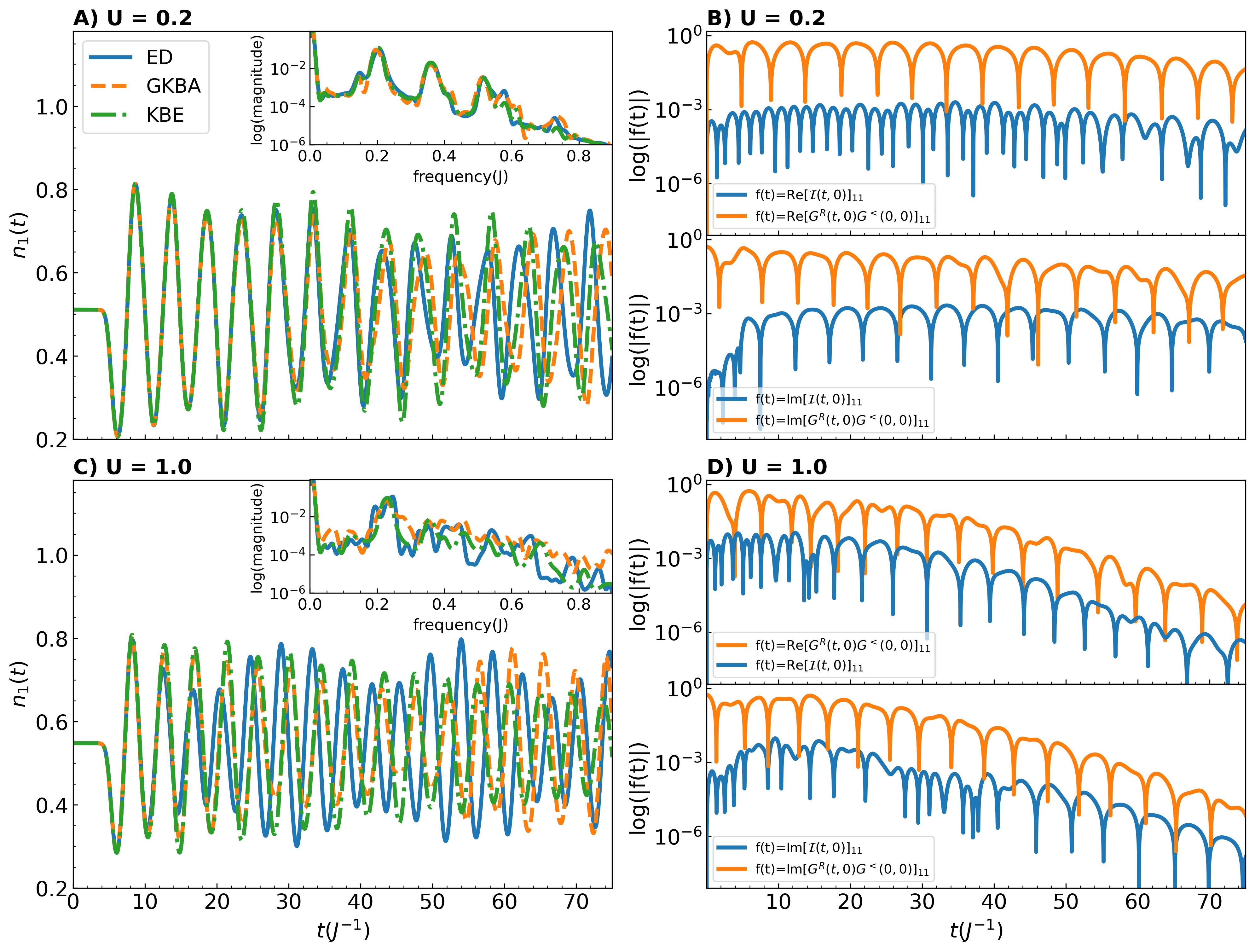}
    \caption{Comparison between HF-GKBA, KBE and exact dynamics in a 4-site model with nearest neighbor hopping and a long wavelength excitation. The magnitude of the excitation is $E=0.5J$. Panels A and C show a comparison of the time-dependent density on the first site of a 4-site chain for $U = 0.2J$ and $U = 1.0J$ respectively.  Panels B and D show the magnitudes of the integral and non-integral terms in equation \eqref{eq:G_lss_full} along the time trajectory $[0,t]$ in the same systems. Inset: Frequency spectrum for the trajectories shown plotted on a semi-log scale .}
    \label{fig:Dipole_kbe_vs_gkba}
\end{figure*}

First, we analyze the results produced in the long wavelength limit.  In panels A and C of Fig. \ref{fig:Dipole_kbe_vs_gkba}, we show the dynamics of the time-dependent site density for the model described by equations \eqref{eq:MB_ham}-\eqref{eq:single_particle} and \eqref{eq:dipole} with $N=4$. The insets show the Fourier spectrum of the corresponding trajectory.  We show dynamics for both weak ($U = 0.2J)$ and strong $(U=1.0J)$ interaction strengths to demonstrate the change of the approximate methods with increased coupling.  Trajectories for the other parameter settings can be seen in Figs. S2-S5 of the supplementary \cite{supp}.  We note that for strong interactions ($U\geq 1.0J$) and strong excitation magnitude ($E=1.0J)$ some of the HF-GKBA simulations were unstable. The reasons for this instability are outlined in \cite{Joost_2022}, and are not due to the error in the numerical propagation scheme, i.e., it will not be fixed by, for example, using a smaller time step in the propagation. In these parameter regimes, we also find the full KBEs are usually stable but often suffer from damping and tend to reach a spurious steady state.  Panels B and D show a selected trajectory for the real and imaginary components of the integral and non-integral terms in equation \eqref{eq:G_lss_full}.

For the long wavelength limit, we look at the model with nearest neighbor hopping, corresponding to $\alpha=\infty$ in equation \eqref{eq:single_particle}.  We believe the KBE solution for the model with long-range hopping $(\alpha=0.7)$ suffers from artificial damping\cite{von_Friesen_2009} and so is not discussed here. However, we note the level of agreement between the HF-GBKA and the exact result is similar for both models.  A plot similar to that shown in Fig. \ref{fig:Dipole_kbe_vs_gkba} is shown in Fig. S1 of the supplementary for the model with long-range hopping.  We have also provided additional figures for the dynamics of the 4-site model with the long wavelength excitation in Figs. S2-S5 and S12. Both models behave characteristically the same. However, the long-range hopping introduces additional frequency peaks that make the dynamics more complicated.    

For the weakly interacting case in panel A of Fig. \ref{fig:Dipole_kbe_vs_gkba} we see both methods capture qualitatively the induced dynamics. In the inset, we see both methods produce very similar frequency spectra.  However, there are differences at various points, leading to a slow deviation growing between the exact and approximate methods. The separation between the exact result and the HF-GKBA grows slightly faster than for the KBEs. Since the HF-GKBA approximates the KBEs, this can be expected.  

In the case of strong interactions, we see the KBEs again slightly improve upon the oscillation frequencies of the time dependent density. Inspecting the trajectory, the difference between the KBEs and the HF-GKBA isn't significantly distinct from the case of weak interactions.  This is an important observation since already, before the approximations leading to HF-GKBA break down, we see a significant difference between the exact dynamics and the dynamics produced by either approximate method.  In the frequency spectrum we see both methods have a much more difficult time following the exact result for strong interactions.  Both approaches capture the position of the main peak well, while the smaller ones are matched quite poorly.  The KBE solution follows the frequency spectrum better, as is reflected in the dynamics, but still misses much of the finer spectral details.

We now turn to panels B and D of Fig. \ref{fig:Dipole_kbe_vs_gkba} where we plot the integral and non-integral terms of equation \eqref{eq:G_lss_full} extracted \emph{exactly} from the full KBE calculation. These integrals can be thought of as correcting the approximate ansatz and so the relative magnitude of the integral and non-integral terms can give us insight to how significant a correction the KBEs offer to the GKBA.To avoid showing all the matrix components we illustrate only the first entry of each and note that the remaining terms have similar behavior and relative magnitudes.  We see for both the weak and strongly interacting case the terms that are kept in the GKBA are orders of magnitude larger than those that are neglected.  Viewing the blue curves in panels B and D of Fig. \ref{fig:Dipole_kbe_vs_gkba} as corrections to the GKBA, we see they are quite insignificant which helps explain the ability of the GKBA to perform extremely similarly to the KBE in these scenarios.  Going from $U = 0.2J$ to $U = 1.0J$ we see very little change in the relative magnitude of the contributions.  This is reflected in the dynamics, where we see a similarly small level of deviation between HF-GKBA and KBE in both panels A and C of Fig. \ref{fig:Dipole_kbe_vs_gkba}.

\begin{figure*}
    \centering
\includegraphics[width=\textwidth]{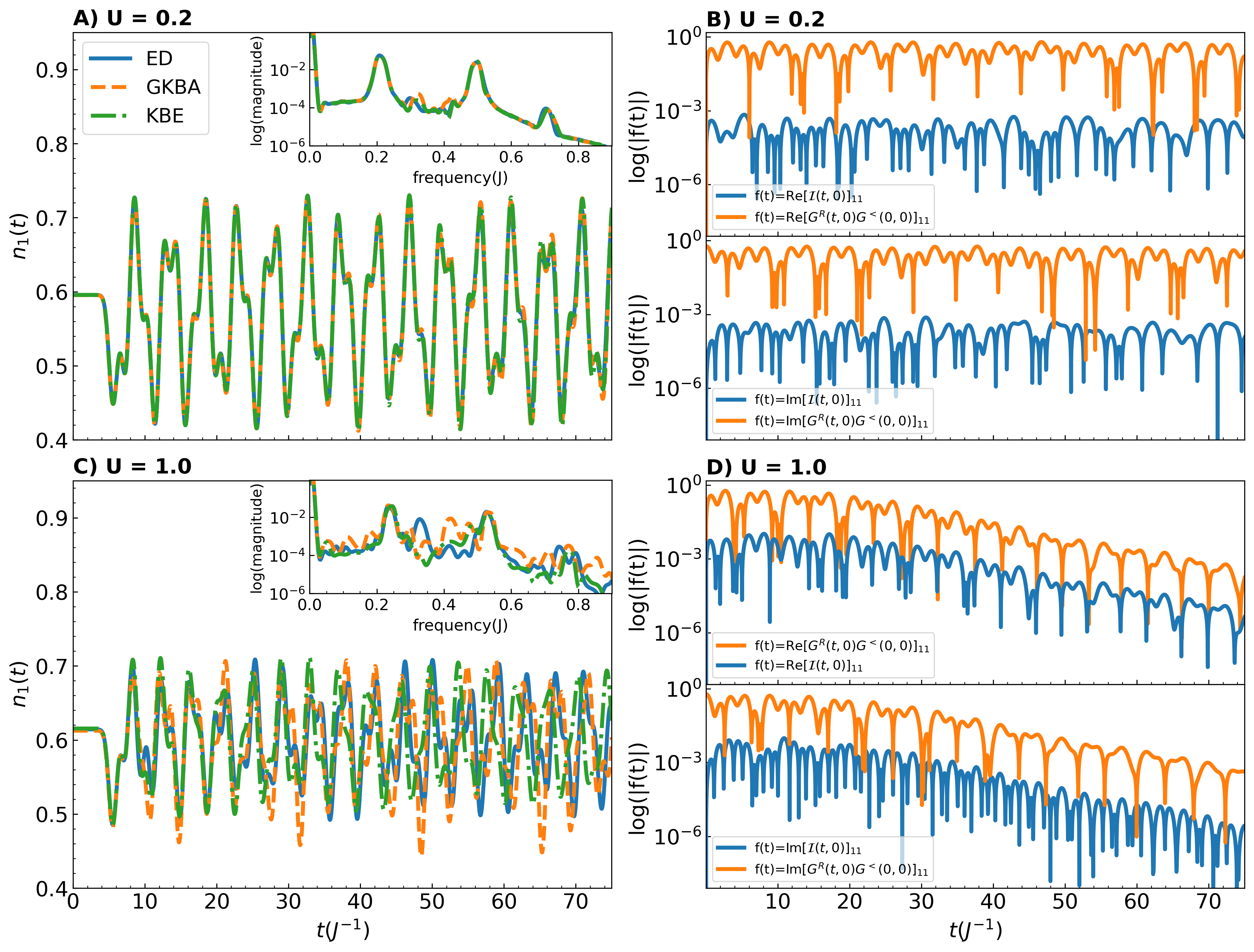}
    \caption{Comparison between HF-GKBA, KBE and exact dynamics in a 4-site model with exponentially decaying long-range hopping and a short wavelength excitation. The magnitude of the excitation is $E=0.5J$. Panels A and C show a comparison of the time-dependent density on the first site of a 4-site chain for $U = 0.2J$ and $U = 1.0J$ respectively.  Panels B and D show the magnitudes of the integral and non-integral terms in equation \eqref{eq:G_lss_full} along the time trajectory $[0,t]$ in the same systems. Inset: Frequency spectrum for the trajectories shown plotted on a semi-log scale.}
    \label{fig:Xray_kbe_vs_gkba}
\end{figure*}
\subsection{Short wavelength limit}\label{sec:short_wavelength}
In this section, we present the results for the four site model with the short wavelength excitation given by equation \eqref{eq:X-ray}. In Fig. \ref{fig:Xray_kbe_vs_gkba} we show the same quantities as in Fig. \ref{fig:Dipole_kbe_vs_gkba}, now for the model which includes long-range inter-site hopping corresponding to $\alpha=0.7$ in equation \eqref{eq:single_particle}.  In the case of nearest neighbor hopping ($\alpha=\infty)$ the dynamics are dominated by frequencies around a single value, making the trajectory much easier to capture.  In contrast, there are at least two dominant spectral peaks when the long-range hopping is included, which is seen clearly in the insets of Fig. \ref{fig:Xray_kbe_vs_gkba}. This makes the dynamics more complicated and more interesting to present here.   As in the case of the long wavelength limit, we include all the trajectories generated for the short wavelength limit in Figs. S7-S10 of the supplementary, a figure similar to Fig. \ref{fig:Xray_kbe_vs_gkba} in Fig. S6 and trajectories up to $125J^{-1}$ for a selection of $U$ values in the different models in Fig. S13 of the supplementary \cite{supp}

For the weakly interacting case ($U=0.2J$) we see excellent agreement between all three methods.  In the strongly interacting case ($U=1.0J)$, as is expected, the agreement worsens but we see still see a relatively good match between the approximate and exact dynamics.  Compared to the long wavelength limit, the oscillation frequencies agree more faithfully.  This is clear looking at both the trajectory and the frequency spectra in Figs. \ref{fig:Dipole_kbe_vs_gkba} and \ref{fig:Xray_kbe_vs_gkba}.  In contrast to panels A and C of Fig. \ref{fig:Dipole_kbe_vs_gkba}, the dynamics in panels A and C of Fig. \ref{fig:Xray_kbe_vs_gkba} do not suffer from a phase shift developing between the approximate and exact results as the time evolution proceeds.

Looking at panels B and D of Fig. \ref{fig:Xray_kbe_vs_gkba} we see a similar trend as in the long wavelength case.  That is, the terms that are kept in the GKBA are orders of magnitude larger than those terms that are dropped.  Thus the correction that keeping these integrals will give to the HF-GKBA will generally be small compared to the term it is correcting.  As with the short wavelength excitation, moving from $U=0.2J$ to $U=1.0J$ we see a slight increase in the integral terms compared to the non-integral terms.  Clearly, for $U = 1.0J$ there is a more pronounced difference between the KBEs and the  HF-GKBA. However, looking at the trajectories a definitive statement about the improvement of the KBEs over the HF-GKBA is harder to make.  This is a similar observation seen in the long wavelength limit, that despite the KBEs and the HF-GKBA differing more significantly, it is not so clear that the KBEs offer significant improvement over the HF-GKBA. In frequency space, the improvement is slightly clearer as the KBEs follows the exact frequency spectrum closer than the HF-GKBA. The largest difference being that the HF-GKBA overestimates a peak around $f=0.45J$.  Both methods capture the main frequency peaks which help explains why neither method seems significantly better in the time domain.

\subsection{Further analysis of the reliability of the GKBA\label{sec:further_analysis}}
The conclusions were so far drawn from mathematical analysis and numerical testing on a 4-site model at various conditions. In this section, we further demonstrate the validity of these observations by considering larger systems ($N=8$) and different fillings ($\bar{n}=\frac12,\frac14)$, which directly influence the magnitude of certain components of the self energy, as discussed below.

\begin{figure*}
    \centering
    \includegraphics[width=\textwidth]{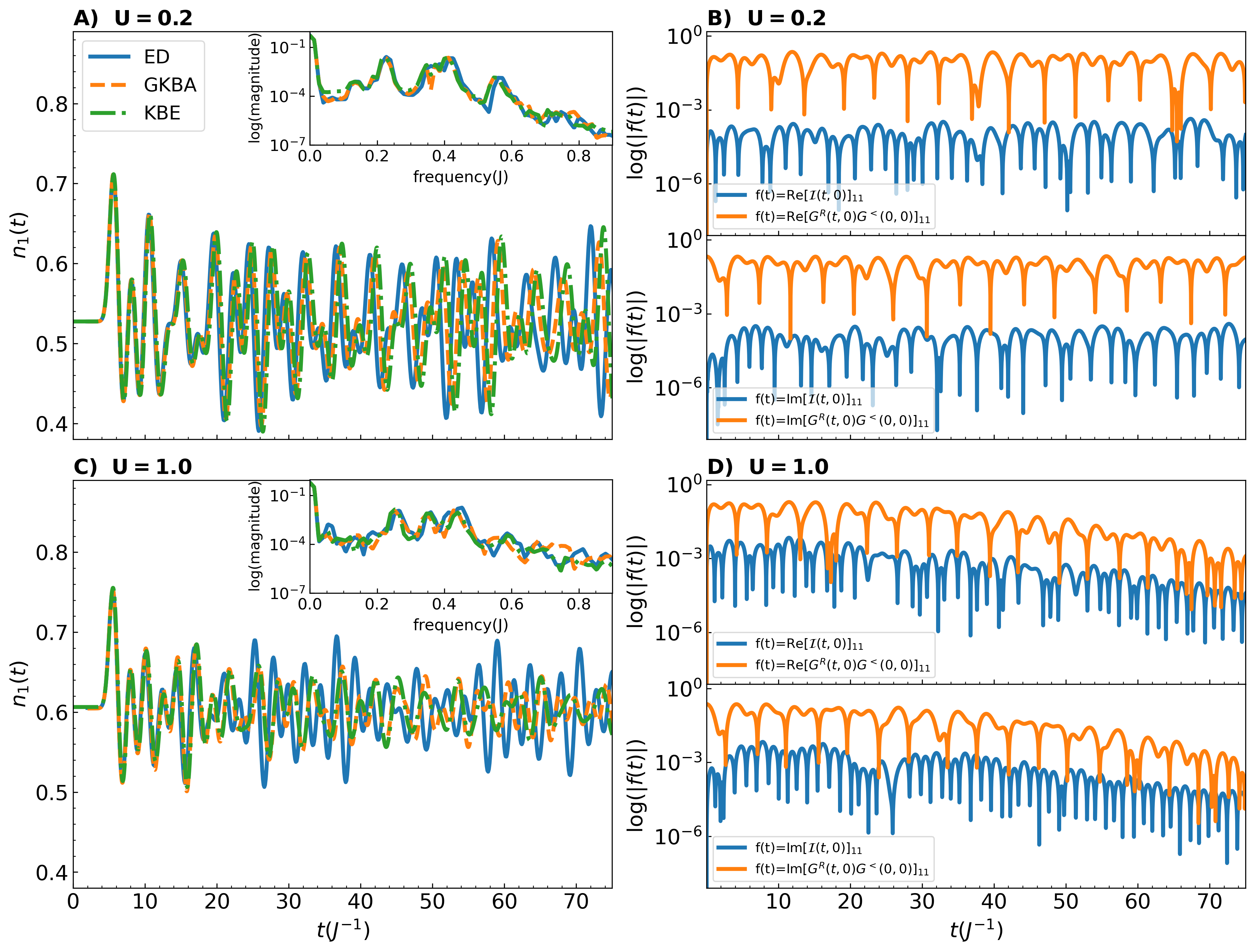}
    \caption{Comparison between HF-GKBA, KBE and exact dynamics in an 8-site model at half filling with nearest neighbor hopping and the short wavelength excitation. The magnitude of the excitation is $E=0.5J$. Panels A and C show a comparison of the time-dependent density on the first site of a 8-site chain for $U = 0.2J$ and $U = 1.0J$ respectively.  Panels B and D show the magnitudes of the integral and non-integral terms in equation \eqref{eq:G_lss_full} along the time trajectory $[0,t]$ in the same systems. Inset: Frequency spectrum for the trajectories shown plotted on a semi-log scale.}
    \label{fig:Xray_kbe_vs_gkba_8}
\end{figure*}

\begin{figure*}
    \centering
    \includegraphics[width=\textwidth]{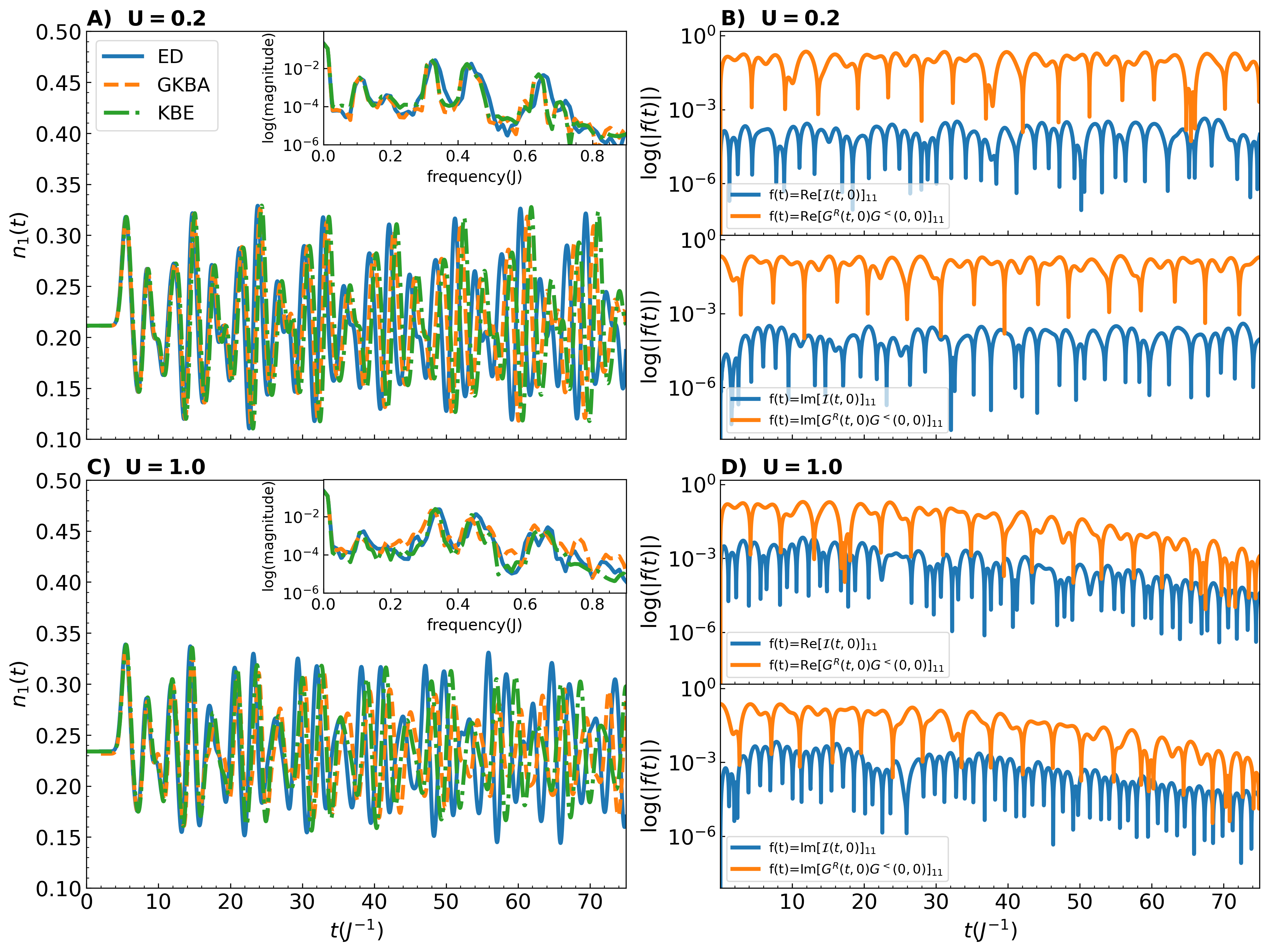}
    \caption{Comparison between HF-GKBA, KBE and exact dynamics in an 8-site model at quarter filling with nearest neighbor hopping and the short wavelength excitation. The magnitude of the excitation is $E=0.5J$. Panels A and C show a comparison of the time-dependent density on the first site of a 8-site chain for $U = 0.2J$ and $U = 1.0J$ respectively.  Panels B and D show the magnitudes of the integral and non-integral terms in equation \eqref{eq:G_lss_full} along the time trajectory $[0,t]$ in the same systems. Inset: Frequency spectrum for the trajectories shown plotted on a semi-log scale.}
    \label{fig:Xray_kbe_vs_gkba_8_quarter}
\end{figure*}

For this additional analysis, we choose the short wavelength excitation as this gave more interesting varied dynamics than the long wavelength excitation.  In Fig. \ref{fig:Xray_kbe_vs_gkba_8}, we show similar plots to those shown in Figs. \ref{fig:Dipole_kbe_vs_gkba} and \ref{fig:Xray_kbe_vs_gkba}.  The simulations shown in Fig. \ref{fig:Xray_kbe_vs_gkba_8} correspond to the 8-site model at half filling with nearest neighbor hopping($\alpha=\infty)$ and in Fig. \ref{fig:Xray_kbe_vs_gkba_8_quarter} we show results for the model with long range hopping($\alpha = 0.7$) at quarter filling.

Starting with Fig \ref{fig:Xray_kbe_vs_gkba_8}, for the weakly interacting case ($U = 0.2J$), as in the previous examples, we see very good agreement between all three methods.  Similar to the long wavelength case in Fig. \ref{fig:Dipole_kbe_vs_gkba}, there is a slight mismatch between some of the frequencies produced by the KBEs and GKBA when compared to the exact result.  Both treatments, GKBA and KBE, capture fine details of the exact result very well, yet a slowly growing shift in the signals is observed as the propagation time increases.  However, we note that this frequency shift seems to affect both schemes to a similar degree and so does \textit{not} mark a significant failure of the GKBA beyond the failings of the KBEs.

For the strongly interacting case, the behavior corresponds to that for the 4-site examples in Figs.~\ref{fig:Dipole_kbe_vs_gkba} and \ref{fig:Xray_kbe_vs_gkba}.  Specifically, we see that although the agreement between the approximate and exact methods worsens going from $U = 0.2J$ to $U = 1.0J$, the KBE and GKBA results remain very close. So, as with our previous results, it appears that the further approximations made in the GKBA remain well justified even when the self energy approximations made in the full KBEs lead to significant deviation from the exact result.

Panels B and D of Fig. \ref{fig:Xray_kbe_vs_gkba_8} also tell a very similar story to the previous results.  There is orders of magnitude difference between the integral and non-integral terms of equation \eqref{eq:G_lss_full}.  This difference diminishes slightly in the strongly interacting case, however, the memory integrals still appear to add only a slight correction to the GKBA.  This is observed in the closeness of the GKBA and KBE curves in both panels A and B of Fig. \ref{fig:Xray_kbe_vs_gkba_8}. The insets showing frequency spectra of the two main plots also have similar behavior to those shown in previous subsections.  That is, both methods capture the major frequencies well for both weak and strong interactions. As before, for strong interactions, the KBEs perform slightly better. For instance, the features around $f=0.5J - 0.6J$ in Fig. \ref{fig:Xray_kbe_vs_gkba_8} panel C are matched better by the KBEs than the GKBA.  However, these differences have only a small impact on the overall time evolution trajectory.  Thus, these results for a larger system further back up the observations made in the previous subsections.

We now discuss the result for the 8-site system away from the half-filling (namely at $\bar n = \frac{1}{4}$) and with long-range hopping (Fig. \ref{fig:Xray_kbe_vs_gkba_8_quarter}).  This is an important test case as the scaling bound placed on the integrand of equation \eqref{eq:G_lss_full} depends not only on the interaction parameter ($w_{ijkl} \sim U$) but also on the density, as is demonstrated in Appendices \ref{sec:app_selfE} and \ref{sec:app_selfE_NE}.  The KBE and GKBA solutions agree well in both panels A and C of Fig. \ref{fig:Xray_kbe_vs_gkba_8_quarter}. The relative magnitude of the memory integral terms (panels B and D) demonstrates that the quarter-filling case performs similarly to the half-filling case.  Incidentally, the KBE and GKBA results even seem to match the magnitude and phase of oscillations slightly better than the half-filling case at strong interactions (panel C in Fig.~\ref{fig:Xray_kbe_vs_gkba_8}). As with our previous examples, in frequency space we see the GKBA performs worse than the KBEs for some of the higher frequency peaks, e.g. around $f = 0.55J$ and $f = 0.75J$, however this is only weakly reflected in the overall dynamics.

\section{Discussion}\label{sec:discussion}

We will now discuss the implications of these results for the applicability of the HF-GKBA. Figs. \ref{fig:Dipole_kbe_vs_gkba}, \ref{fig:Xray_kbe_vs_gkba}, \ref{fig:Xray_kbe_vs_gkba_8} and \ref{fig:Xray_kbe_vs_gkba_8_quarter} as well as Figs. S1-S10 in the supplemental\cite{supp} show that both the KBEs and the HF-GKBA perform well at capturing the dynamics of the system under excitations by an external electric field.  For weak interactions, the KBEs offers only a marginal improvement over the dynamics produced by the HF-GKBA. This is partially explained by the comparison in the magnitude of the integral and non-integral terms in panels B and D of Figs. \ref{fig:Dipole_kbe_vs_gkba}, \ref{fig:Xray_kbe_vs_gkba}, \ref{fig:Xray_kbe_vs_gkba_8} and \ref{fig:Xray_kbe_vs_gkba_8_quarter}.  When one moves to stronger interactions(up to $U=1.0J$) both methods still capture the induced dynamics qualitatively. However, as is to be expected, a clear worsening of the approximate dynamics is seen.  We also see a larger difference between the HF-GKBA and the full KBEs in panel C of Figs. \ref{fig:Dipole_kbe_vs_gkba}, \ref{fig:Xray_kbe_vs_gkba}, \ref{fig:Xray_kbe_vs_gkba_8} and \ref{fig:Xray_kbe_vs_gkba_8_quarter} compared to panel A. This is reflected somewhat in the change in the relative magnitude of the integral and non-integral terms of equation \eqref{eq:G_lss_full} in panel D.  In these cases, it is not obvious that the full KBEs offer an improvement over the HF-GKBA in reproducing the density matrix dynamics. For these stronger interactions, issues arising with the self energy expansion will dominate thus making the memory effects a secondary issue.

Comparing the magnitude of the integral and non-integral terms in equation \eqref{eq:G_lss_full} gives us a measure of how HF-GKBA and the KBEs will differ from one another.  This is simply because the integral terms can be thought to correct the HF-GKBA and so their relative magnitude shows how significant a correction there is.  This measure is meant to quantitatively demonstrate that the integrals can be rather small, even for strong interactions ($U=1.0J)$.  The arguments in section \ref{sec:markovian_analysis} and the scaling bound provided in appendices \ref{sec:app_selfE} and \ref{sec:app_selfE_NE} help explain the small size of these integrals.  Furthermore, here, the integral terms do not increase but rather oscillate with an amplitude much smaller than the non-integral terms.  This likely adds to the insignificant contribution of the memory integrals since, in addition to the decaying nature of the self energy, the memory integral will also have a harder time building up in magnitude due to the constant oscillation of the integrand.  

The validity of the GKBA was tested in an alternate way in \cite{Kwong_1998} by using the retarded and advanced propagators from a full KBE calculation in the GKBA in equation \eqref{eq:HF-GKBA}.  This gave strong agreement between the GKBA and full KBEs; in line with our results here, this suggests the unimportance of the neglected memory integrals. Here, we go beyond this test and perform a detailed investigation of the neglected terms and how they scale, with regards to parameters of the system, to concretely explain the often seen similarity between the GKBA and KBEs, particularly in microscopic and finite systems.  The in-depth analysis of the neglected terms explains the reason behind the quality of the ansatz rigorously using the theoretical bounds for the magnitudes of the non-local memory terms.

The tests we have performed in the 8-site system help to strengthen the conclusions we draw.  Going to this larger system helps to diminish small system size effects.  The results for the 4 and 8 site models tell a very similar story and give the promising suggestion that our arguments are independent of the system size.  Furthermore, it has been argued\cite{Hermanns_2014}, from results in similarly sized systems, that the GKBA systematically improves with the number of sites; hence,  we believe that our conclusions can be extrapolated to systems even larger than those tested here.

The final test we performed looked at the effect of the filling factor on our arguments.  The bound placed on the integrand in equation \eqref{eq:G_lss_full} depends on the density and the interaction strength.  To the leading order, the density dependence of the bound is cubic in the particle or hole density, depending on which is larger in magnitude (more details are provided in Appendices \ref{sec:app_selfE} and \ref{sec:app_selfE_NE}).  Going to a system with quarter filling increases the average hole density and thus should increase the value of this bound and the integrand.  We observed little change from the previous cases tested and conclude that densities very far from half filling will be needed to significantly impact the scaling of the integrand. As seen in Fig. \ref{fig:Xray_kbe_vs_gkba_8_quarter}, even for a relatively low filling factor, the arguments we put forth seem to hold well, and the KBEs and GKBA produce results with only minor differences from one another. 

We note that although the approximate dynamics in panel C of Figs. \ref{fig:Dipole_kbe_vs_gkba}, \ref{fig:Xray_kbe_vs_gkba}, \ref{fig:Xray_kbe_vs_gkba_8} and \ref{fig:Xray_kbe_vs_gkba_8_quarter} look quite different from the exact result, the position of the main spectral peaks are still captured and so the trajectories still give useful information of the non-equilibrium properties of the system.    

We will now discuss the role of self energy in these calculations.  The most obvious strategy for improving either the full KBEs or the HF-GKBA is through the form of $\Sigma$.  In the case of the KBEs this is often not done, and while existing, KBE simulations with self energies beyond second Born are uncommon\cite{Attaccalite_2011,Schuler_2020,Schlunzen_2016,Schlunzen_2017,Schlunzen_2016_2}.  This is due to the extra overhead incurred at each time step. In the case of the HF-GKBA, the much more favorable numerical scaling and lower computational cost make it  significantly easier to implement improved $\Sigma$. $GW$ approximation is a concrete example:  Existing implementations of the full KBEs with $GW$ self energy are severely limited by system size and length of the time evolution.  This makes these implementations impractical for studying extended systems where the $GW$ self energy is mainly used and where it offers the most significant improvements\cite{Reining_2018}.  In contrast, HF-GKBA implementations with the $GW$ approximation can be used in much larger systems and over longer times than the full KBEs(even with the second Born self energy).  The use of a more advanced self energy can lead to a dramatic improvement in results both in and out of equilibrium\cite{Joost_2022,Attaccalite_2011,Mejuto_2022,Pavlyukh_2022,Pavlyukh_2021,Pavlyukh_2022_2}. 

\section{Conclusions}
Throughout this paper, we have made use of the commonly applied HF-GKBA.  The arguments put forth in section \ref{sec:markovian_analysis} only looked at the approximation leading to the GKBA.  The further approximations leading to the HF-GKBA are similar in nature and involve neglecting memory integrals in the equation of motion for the advanced and retarded propagators.  For this reason, we expect similar arguments as given in section \ref{sec:markovian_analysis} to hold for these approximations.  We also believe the primary effect of this approximation will present itself in the spectral properties since these are directly related to the retarded/advanced GF. An interesting future direction could be the implementation of improved off-diagonal reconstructions, using static self energies to retain the computational advantage of the HF-GKBA\cite{Lorke_2006,Bonitz_2019}.  This will improve the spectral properties and allow for improved theoretical studies of time resolved ARPES experiments and phenomena such as band gap renormalization\cite{Freericks_2009,Spataru_2004}.

In this paper, we have given arguments and numerical evidence in experimentally relevant setups, that the effects neglected in deriving the HF-GKBA offer, in many cases, only a small correction to approximate dynamics generated.  Furthermore, we see in many scenarios under which the HF-GKBA does begin to deviate from exact results, the full KBEs also deviate to a similar degree and no longer give an obvious improvement to the HF-GKBA.  We believe that the ability to practically implement more complex self energies in the HF-GKBA is far more important for capturing the particle dynamics than including the neglected memory effects but relying, for example, on the second Born self energy.  This is because the neglected memory effects seem important, primarily  in the case of strong interactions, where even equilibrium properties are difficult to predict with diagrammatic methods.  Conversely, in regimes where equilibrium properties are captured closely by diagrammatic methods, the HF-GKBA seems to generally perform quite well for non-equilibrium properties. We predict that improvements to equilibrium properties from using an improved self energy approximation will carry over to non-equilibrium properties also.  Thus it is likely to be more fruitful to use a self energy approximation such as $GW$ or beyond\cite{Mejuto_2022,Mejuto_2021} within the HF-GKBA, rather than use the full KBEs with simple static self energy such as the second Born. This is especially true in extended systems where not only do we need to be able to simulate large systems, but also the screening effects that $GW$ accounts for become more important to include\cite{Reining_2018}.

Finally, we note that the arguments put forth in this paper hope to elucidate some of the reasons why the HF-GKBA often performs extremely well compared to exact results and the KBEs in finite systems\cite{Hermanns_2014,Schlunzen_2017}. The KBEs and HF-GKBA have been applied in the study of macroscopic systems, e.g., correlated electron gases and dense quantum plasmas, with indications that the improvements offered by the full KBEs are more significant than in finite systems\cite{Kwong_2000,Kremp_1999,Bonitz_1996,Banyai_1998,Haberland_2001,Kwong_1998} and lack spurious damping of the KBEs. Even there, however, KBE still suffers from unfavorable cubic scaling. Our observations and claims made here for microscopic and finite systems show that the GKBA however performs very similarly to KBE.

In practice, HF-GKBA emerges as a useful tool to study non-equilibrium dynamics in such systems as it has many selling points:  linear scaling in the number of time steps, practical ability to use a range of non-static self energies\cite{Joost_2022,Joost_2020}, the ability to perform calculations in large extended systems\cite{Perfetto_2022} and in many scenarios the ability to give accurate results\cite{Schlunzen_2017}.  All of these points make it an extremely practical scheme to take NEGF theory beyond the simulation of small lattice models and atomic/molecular systems and into the realm of ab initio studies of real extended systems, previously only accessible through methods such as time-dependent Hartree-Fock or time-dependent density functional theory.

\section*{Acknowledgements}
This material is based upon work supported by the U.S. Department of Energy, Office of Science, Office of Advanced Scientific Computing Research and Office of Basic Energy Sciences, Scientific Discovery through Advanced Computing (SciDAC) program under Award Number DE-SC0022198.  This research used resources of the National Energy Research Scientific Computing Center, a DOE Office of Science User Facility supported by the Office of Science of the U.S. Department of Energy under Contract No. DE-AC02-05CH11231 using NERSC award BES-ERCAP0020089.
\appendix
\section{Scaling limit for the magnitude  of the self energy:time independent case}\label{sec:app_selfE}
We provide a simple mathematical proof for the $O(w^2c^3)$ scaling limit for the magnitude of the second Born self energy \eqref{2ndB_selfE}. Consider a general nonequilibrium case where a time-independent fermionic system described by a Hamiltonian $\H$ starts from initial state $\rho=\sum_n\rho_n|n\rangle\langle n|$, where $|n\rangle$ is the eigenstate of $\mathcal{H}$ and $\hat\rho_{ii}=\sum_n\rho_n\langle n|c_{i}^{\dagger}c_i| n\rangle $ satisfying $0<\hat\rho_{ii}\leq 1$. Here $\hat\rho_{ii}$ is the $i$-th diagonal element of the initial density matrix and for the sake of simplicity we dropped the spin indices. In this case, the Green's function can be written as follows
\begin{align*}
G^<_{ij}(t,t')&=i\sum_n\rho_n\langle n|c_{i}^{\dagger}(t')c_{j}(t)|n\rangle\\
&=i\sum_n\rho_n\langle n|
e^{it'\H}c_{i}^{\dagger}e^{i(t-t')\H}c_{j}e^{-it\H}|n\rangle\\
&=i\sum_{nn'}
\rho_n\langle n|
c_{i}^{\dagger}|n'\rangle\langle n'|c_{j}|n\rangle e^{i(t-t')(E_{n'}-E_{n})}\\
G^>_{ij}(t,t')&=-i\sum_{nn'}\rho_n\langle n|c_i|n'\rangle\langle n'|c_j^{\dagger}|n\rangle e^{i(t-t')(E_n-E_{n'})}
\end{align*}
Using triangle inequality, Cauchy-Schwarz inequality and $\sqrt{ab}\leq \frac{1}{2}(a+b)$, we obtain
\begin{align*}
|G^<_{ij}(t,t')|&\leq 
\sum_{n}\rho_n \sum_{n'}|\langle n|c_i^{\dagger}|n'\rangle||\langle n'|c_j|n\rangle|\\
&\leq \sum_{n}\rho_n\bigg[
\underbrace{\sum_{n'}|\langle n|c_i^{\dagger}|n'\rangle|^2
}_{\langle n|c_i^{\dagger}c_i|n\rangle}
\bigg]^{1/2}
\bigg[
\underbrace{
\sum_{n'}|\langle n'|c_j|n\rangle|^2}
_{\langle n|c_j^{\dagger}c_j|n\rangle}
\bigg]^{1/2}\\
&\leq
\frac{1}{2}\sum_{n}\rho_n\langle n|c_i^{\dagger}c_i|n\rangle+\frac{1}{2}\sum_{n}\rho_n\langle n|c_j^{\dagger}c_j|n\rangle\\
&=\frac{1}{2}(\hat\rho_{ii}+\hat\rho_{jj})=c^<_{ij}\leq 1
\end{align*}
 Similarly, for the greater Green's function, we have 
\begin{align*}
|G^>_{ij}(t,t')|\leq 
\frac{1}{2}[(1-\hat\rho_{ii})+(1-\hat\rho_{jj})]=c^>_{ij}\leq1
\end{align*}
Choosing $0<c=\max_{ij}\{c_{ij}^<,c_{ij}^>\}\leq 1$ and then using the definition of the second Born self energy \eqref{2ndB_selfE}, we could obtain the following estimate:
\begin{align*}
    |\Sigma^{\lessgtr}_\mathrm{SB}(t,t')|\sim w^2 |G^\lessgtr(t,t') G^\lessgtr(t,t') G^\gtrless(t',t)|\sim O(w^2c^3)
\end{align*}
where $|\cdot|$ can be specified to be any matrix norm. We note that the $c$ parameter only depends on the initial density in the system. Although in principle $c$ can be equal to one, in practice in the systems we study the initial state always has $\rho_{ii} < 1$ and so $c<1$.

\section{Scaling limit for the magnitude  of the self energy: non-equilibrium case}\label{sec:app_selfE_NE}
Here we extend the proof in the previous section to the non-equilibrium case. We assume the time evolution to be unitary and determined by a time-dependent fermionic Hamiltonian $\H(t)$. Starting from the initial state $\rho=\sum_n\rho_n|n\rangle\langle n|$, where $|n\rangle$ is the eigenstate of $\mathcal{H}(0)$.  We now can write the following,
\begin{align*}
G^<_{ij}(t,t')&=i\sum_n\rho_n\langle n|c_{i}^{\dagger}(t')c_{j}(t)|n\rangle,\\
&=i\sum_n\rho_n\langle n |U(0,t')c^\dagger_iU(t,t')c_jU(t,0)|n\rangle,\\
&=i\sum_{n,n_1} \rho_n \langle \Tilde{n} | c_i^\dagger |n_1\rangle\langle n_1|U(t,t')c_j|\Tilde{n}'\rangle,\\
&=i\sum_{n,n_1} \rho_n \langle \Tilde{n} | c_i^\dagger |n_1 \rangle \langle n_1|c_j|\Tilde{n}'\rangle e^{i\Tilde{E}_{n_1}(t,t') }.
\end{align*}
We make use of the fact that the unitary operator can be written as the complex exponential of a Hermitian operator and choosing the state $|n_1\rangle$ so that,
\begin{align*}
    \Tilde{\mathcal{H}}(t,t') | n_1\rangle = \Tilde{E}_{n_1}(t,t')|n_1\rangle
\end{align*}
where $ \Tilde{\mathcal{H}}(t,t')$ is the Hermitian matrix such that,
\begin{align*}
    U(t,t')=e^{-i\Tilde{\mathcal{H}}(t,t')}
\end{align*}
We also have defined two auxiliary states as,
\begin{align*}
    |\Tilde{n}\rangle &= U(t',0)|n\rangle\\
    |\Tilde{n}'\rangle &= U(t,0)|n\rangle\\
\end{align*}
Using the triangle inequality, the Cauchy-Schwarz inequality and $\sqrt{ab}\leq \frac{1}{2}(a+b)$, as in appendix \ref{sec:app_selfE} we obtain
\begin{align*}
|G^<_{ij}(t,t')|&\leq 
\sum_{n}\rho_n \sum_{n_1}|\langle n|c_i^{\dagger}|n_1\rangle||\langle n_1|c_j|n\rangle|\\
&\leq \sum_{n}\rho_n\bigg[
\underbrace{\sum_{n_1}|\langle \Tilde{n}|c_i^{\dagger}|n_1\rangle|^2
}_{\langle \Tilde{n}|c_i^{\dagger}c_i|\Tilde{n}\rangle}
\bigg]^{1/2}
\bigg[
\underbrace{
\sum_{n_1}|\langle n_1|c_j|\Tilde{n}'\rangle|^2}
_{\langle \Tilde{n}'|c_j^{\dagger}c_j|\Tilde{n}'\rangle}
\bigg]^{1/2}\\
&\leq
\frac{1}{2}\sum_{n}\rho_n\langle \Tilde{n}|c_i^{\dagger}c_i|\Tilde{n}\rangle+\frac{1}{2}\sum_{n}\rho_n\langle \Tilde{n}'|c_j^{\dagger}c_j|\Tilde{n}'\rangle\\
&\leq
\frac{1}{2}\sum_{n}\rho_n+\frac{1}{2}\sum_{n}\rho_n = 1\\
\end{align*}
Here moving from the second last to last line we use the fact that $\langle \Tilde{n}|c_i^{\dagger}c_i|\Tilde{n}\rangle$ is just an occupation number for the state $|\Tilde{n}\rangle$ and so $\langle \Tilde{n}|c_i^{\dagger}c_i|\Tilde{n}\rangle \leq 1$. 
Similarly, for the greater Green's function, we have 
\begin{align*}
|G^>_{ij}(t,t')|
&\leq
\frac{1}{2}\sum_{n}\rho_n\langle \Tilde{n}|c_ic_i^{\dagger}|\Tilde{n}\rangle+\frac{1}{2}\sum_{n}\rho_n\langle \Tilde{n}'|c_jc_j^{\dagger}|\Tilde{n}'\rangle\\
&=
\frac{1}{2}\sum_{n}\rho_n\langle \Tilde{n}|1-c_i^{\dagger}c_i|\Tilde{n}\rangle+\frac{1}{2}\sum_{n}\rho_n\langle \Tilde{n}'|1-c_j^{\dagger}c_j|\Tilde{n}'\rangle\\
&=c_{ij}^{>}\\
&\leq
\frac{1}{2}\sum_{n}\rho_n+\frac{1}{2}\sum_{n}\rho_n = 1.\\
\end{align*}
 We can now use a similar analysis given at the end of appendix \ref{sec:app_selfE} and so the scaling limit $O(\omega^2c^3)$ is valid also for the time-dependent case.   With these results, we conclude that all integrands in \eqref{eq:G_lss_full} scale as of $O(\omega^2c^3)$.    We note now that the $c$ parameter depends on the density at a specific time and can still in principle be equal to one.  However, again, in practice in the systems we study the density is rarely completely localizes on one site and so typically $c < 1$.  Therefore,  for the case when $\omega\sim O(1)$, e.g. panel D in Fig. \ref{fig:Dipole_kbe_vs_gkba} and Fig. \ref{fig:Xray_kbe_vs_gkba}, we still see the relatively large difference between the integral and non-integral terms of equation \eqref{eq:G_lss_full}.

\bibliography{Bib}
\end{document}